\documentclass[twocolumn,showpacs,pre,aps,preprintnumbers,floatfix,amssymb,showkeys]{revtex4}
\usepackage{graphicx}
\begin{document}
\preprint{APS/EU9054}
\title{Fine structure generation in double-diffusive system}
\author{S.B. Kozitskiy}
\email{skozi@poi.dvo.ru}
\affiliation{Laboratory of Wave Phenomena Modeling,
V.I. Il'ichev Pacific Oceanological Institute of the Russian Academy of Sciences,
690041, Baltiyskaya 43, Vladivostok, Russia}
\date{\today}
\begin{abstract}
Double-diffusive convection in a horizontally infinite layer of a
unit height in a large Rayleigh numbers limit is considered. From
linear stability analysis it is shown, that the convection tends
to have a form of travelling tall thin rolls with width about 30
times less than height. Amplitude equations of ABC type for
vertical variations of amplitude of these rolls and mean values of
diffusive components are derived. As a result of its numerical
simulation it is shown, that for a wide variety of parameters
considered ABC system have solutions, known as diffusive chaos,
which can be useful for explanation of fine structure generation
in some important oceanographical systems like thermohaline
staircases.
\end{abstract}



\maketitle

\section{INTRODUCTION}

Double-diffusive or thermohaline convection plays important role
in a heat-mass transfer processes in the ocean~\cite{Turner:1973}.
It also essentially influences on different small scale processes,
like formation of vertical temperature and salinity fine
structure. Such phenomena are not well understood at present day.
There are only few works, devoted to analytical models of fine
structure generation in the
sea~\cite{Grimshaw:1979,Grimshaw:1982,Vor}. But no one of them
essentially considers the role of double diffusion in such
processes. As an exception one can mention
work~\cite{Kerstein:1999}, where double-diffusive step-like fine
structure is simulated by numerical Monte Carlo methods. The
purpose of this article is to develop one more mathematical model
of two dimensional double-diffusive convection in a horizontally
infinite layer, based on system of amplitude equations, describing
formation of vertical fine structure, which in some aspects
resembles actual experimental and observational data. The main
idea of this work consists in combining the result, that in the
limit of large Rayleigh numbers convective cells tend to be narrow
and tall, with constructing of ABC-system of amplitude
equations~\cite{Balmforth:1998} with respect to vertical
coordinate in the case of such cells. So this article includes two
main sections apart from this one. In the section \ref{sec2} from
linear stability problem we determine sizes of the most prominent
cells in large Rayleigh numbers limit. In the section \ref{sec3}
in weakly nonlinear approximation by multi-scale decomposition
technique we derive the ABC-system. Its numerical simulation gives
vertical fine structure via so called diffusive chaos solutions.

The initial equations describe two-dimensional thermohaline
convection in a liquid layer of thickness $h$, bounded by two
infinite plane horizontal boundaries. The liquid moves in a
vertical plane and the motion is described by the stream function
$\psi(t,x,z)$. The horizontal $x$ and vertical $z$ space variables
are used; the time is denoted by $t$. It is assumed, that there
are no distributed sources of heat and salt, and on the upper and
lower boundaries of the area these quantities have constant
values. Hence, basic distribution of temperature and salinity is
linear along the vertical and is not depend on time. The variables
$\theta(t,x,z)$ and $\xi(t,x,z)$ describe variations in the
temperature and salinity about this main distribution. There are
two types of thermohaline convection: the
fingering~\cite{Sorkin:2002}, in which the warmer and more saline
liquid is at the upper boundary of the area, and the diffusive
type, in which the temperature and salinity are greater at the
lower boundary~\cite{Turner:1974}. In this paper we study the
later case.

The governing equations in the Boussinesq approximation in
dimensionless form are a system of nonlinear equations in first
order partial derivatives with respect to time, that depend on four
parameters: the Prandtl number $\sigma$ (usual value - $7.0$), the
Lewis number $\tau$ ($0<\tau<1$, usually $0.01-0.1$), and the
temperature $R_T$ and salinity $R_S$ Rayleigh numbers
\cite{Knobloch:1986,K10}:
\begin{eqnarray} \label{maineq}
& & (\partial_t - \sigma \Delta)\Delta\psi + \sigma (R_S
\partial_x \xi - R_T \partial_x \theta) = J(\Delta\psi,\psi), \nonumber \\
& & (\partial_t - \Delta)\theta - \partial_x\psi = J(\theta,\psi), \\
& & (\partial_t - \tau\Delta)\xi - \partial_x\psi = J(\xi,\psi).
\nonumber
\end{eqnarray}
Here the Jacobian $J(f,g)=\partial_{x}{f}\partial_{z}{g}-
\partial_{x}{g}\partial_{z}{f}$ is introduced. First equation in
(\ref{maineq}) describes liquid particle pulse evolution in terms
of stream function, second and third ones describe temperature and
salt diffusion respectively. The boundary conditions for the
dependent variables are chosen to be zero, which implies that the
temperature and salinity at the boundaries of the area are
constants, the vorticity vanishes at the boundaries, and the
boundaries are impermeable:
\begin{equation} \label{econ} % (2)
\psi = \partial_z^2 \psi = \theta = \xi = 0 \mbox{ on } z=0,\,\,1.
\end{equation}
These boundary conditions are usually called free-slip
conditions because the horizontal velocity component at the
boundary does not vanish.

As a space scale the thickness of the liquid layer $h$ is used. As
a time scale value $t_0={h^{2}}/{\chi}$ is used, where $\chi$ is
the thermal diffusivity of the liquid. Velocity field components
are determined as $v_{z} = ({\chi}/{h}){\partial_{x}}{\psi}$ and
$v_{x} = - ({\chi}/{h}){\partial_{z}}{\psi}$. For temperature $T$
and salinity $S$ we have relations:
\begin{eqnarray*}
&& T(t,x,z)=T_{-}+\delta {T}\left[1-z+\theta(t,x,z)\right], \\
&& S(t,x,z)=S_{-}+\delta {S}\left[1-z+\xi(t,x,z)\right].
\end{eqnarray*}
Here $\delta T=T_{+}-T_{-}$, $\delta S=S_{+}-S_{-}$, where
$T_{+}$, $T_{-}$ and $S_{+}$, $S_{-}$ are the
temperatures and salinities on the lower and upper boundaries of
the region, respectively. The temperature and salinity Rayleigh
numbers can be expressed as follows:
\begin{eqnarray*}
R_T = \frac{{g}{\alpha'}{h^{3}}}{\chi\nu}{\delta}{T}, \qquad R_S =
\frac{{g}{\gamma'}{h^{3}}}{\chi\nu}{\delta}{S},
\end{eqnarray*}
where $g$ is the acceleration of gravity, $\nu$ is the viscosity of
the liquid, $\alpha'$ and $\gamma'$ are the temperature and salinity
coefficients of volume expansions.

\section{FORM OF CONVECTIVE CELLS AT LARGE RAYLEIGH
NUMBERS}\label{sec2}

Consider thermohaline convection in a limit of large $R_S$, which is
true for the most of oceanographically important applications ($R_S
\approx 10^{9} - 10^{12}$). After rescaling of the time $t=(\sigma
R_S)^{-1/2}t'$, and the stream function $\psi=(\sigma
R_S)^{1/2}\psi{'}$, we can rewrite basic system (\ref{maineq}) in a
singularly disturbed form (primes are omitted):
\begin{eqnarray} \label{maineq2} % (3)
& & (\partial_t - \sigma\varepsilon^2\Delta)\Delta\psi +
(\partial_x \xi - (1-N^2) \partial_x\theta) = J(\Delta\psi,\psi),  \nonumber \\
& & (\partial_t - \varepsilon^2\Delta)\theta - \partial_x\psi =
 J(\theta,\psi), \\
& & (\partial_t - \tau\varepsilon^2\Delta)\xi - \partial_x
 \psi = J(\xi,\psi), \nonumber
\end{eqnarray}
Here a small parameter $\varepsilon^4 = 1/\sigma R_S$ and a
buoyancy frequency $N^2 = 1-R_T/R_S$ are introduced. In this
system singular perturbations are present as $\varepsilon^2$
before Laplacians. If we let $\varepsilon=0$, then our system
(\ref{maineq2}) turns into common equations, describing
two-dimensional internal waves with the constant buoyancy
frequency $N$ in the Boussinesq approximation.

For investigating of a linear stability problem for the system
(\ref{maineq2}) with boundary conditions (\ref{econ}) we omit
nonlinear terms in the right part of the system and choose a
solution in a form of normal mode:
\begin{eqnarray}\label{ncm2} % (4)
\psi(x,z,t) &=& A e^{\lambda t - ikx}\sin{n\pi z}, \nonumber \\
\theta(x,z,t) &=& a_T e^{\lambda t - ikx}\sin{n\pi z}, \\
\xi(x,z,t) &=& a_S e^{\lambda t - ikx}\sin{n\pi z}. \nonumber
\end{eqnarray}
where $\lambda$ is an eigen value, describing growth rate of the
mode, $k$ is a horizontal wave number, $n$ is a number of the
mode and $A$ is an amplitude of the mode. After substitution of
the expressions (\ref{ncm2}) into the system (\ref{maineq2}) we
get a system of algebraic equations with solvability condition,
having a form of a third order polynomial with respect to $\lambda$:
\begin{eqnarray}\label{dis2} % (5)
(\lambda+\sigma\varepsilon^2\varkappa^2)(\lambda+\varepsilon^2\varkappa^2)
(\lambda+\tau\varepsilon^2\varkappa^2)&& \nonumber \\
+ \frac{k^2 N^2}{\varkappa^2}(\lambda+\gamma\varepsilon^2\varkappa^2) = 0.
\end{eqnarray}
Here $\varkappa^2=k^2+n^2\pi^2$ is a full wave number and $\gamma$
is a constant: $\gamma=\tau+(1-\tau)/N^2$. Equation (\ref{dis2})
is known as dispersive relation and has three roots, two of which
can be complex conjugates for a sufficiently small value of
$\varepsilon$. In the later case a Hopf bifurcation take place
when at some values of $N$ and $\varepsilon$ real part of the
complex conjugates roots turns to zero. It is true when
\begin{eqnarray*}
\varepsilon^4 &<&
\frac{k^2}{\tau^2\varkappa^6}\left(\frac{1-\tau}{1+\sigma}\right), \\
N^2_{*} &=& \frac{1-\tau}{1+\sigma} -
\varepsilon^4\frac{\varkappa^6}{k^2}
[\sigma+\tau(1+\tau+\sigma)], \\
\omega^2 &=&
\frac{k^2}{\varkappa^2}\left(\frac{1-\tau}{1+\sigma}\right)-
\varepsilon^4\tau^2\varkappa^4.
\end{eqnarray*}
where $\omega={\rm Im}(\lambda)$ is a Hopf frequency.

Because dispersive relation (\ref{dis2}) explicitly contains small
parameter, we can choose one of the complex conjugates roots and
express $\lambda$ in the form of an asymptotic expansion by the
powers of $\varepsilon$:
\[
\lambda = \lambda_0+\varepsilon^2\lambda_1+\varepsilon^4\lambda_2
+\varepsilon^6\lambda_3+\cdots
\]
After substitution of this expression in (\ref{dis2}) we have
for $\lambda_i$:
\begin{eqnarray*}
\lambda_0^2 &=& -({k^2}/{\varkappa^2})N^2 \qquad
\lambda_1=\varkappa^2 F_1, \\
\lambda_2 &=& -({\varkappa^4}/{\lambda_0}) F_2, \qquad
\lambda_3=-{\varkappa^8}/({k^2 N^2}) F_3, \\
F_1 &=& (\gamma-C_1)/2>0, \\
F_2 &=& (3F_1^2+2C_1 F_1+C_2)/2>0, \\
F_3 &=& 4 F_1^3+4 C_1 F_1^2+(C_1^2+C_2)F_1 \\
&& +(\tau+\sigma)(C_1+\tau\sigma)/2>0,
\end{eqnarray*}
\begin{figure}[ptb]
\centering \includegraphics[width=2.3in,angle=-90]{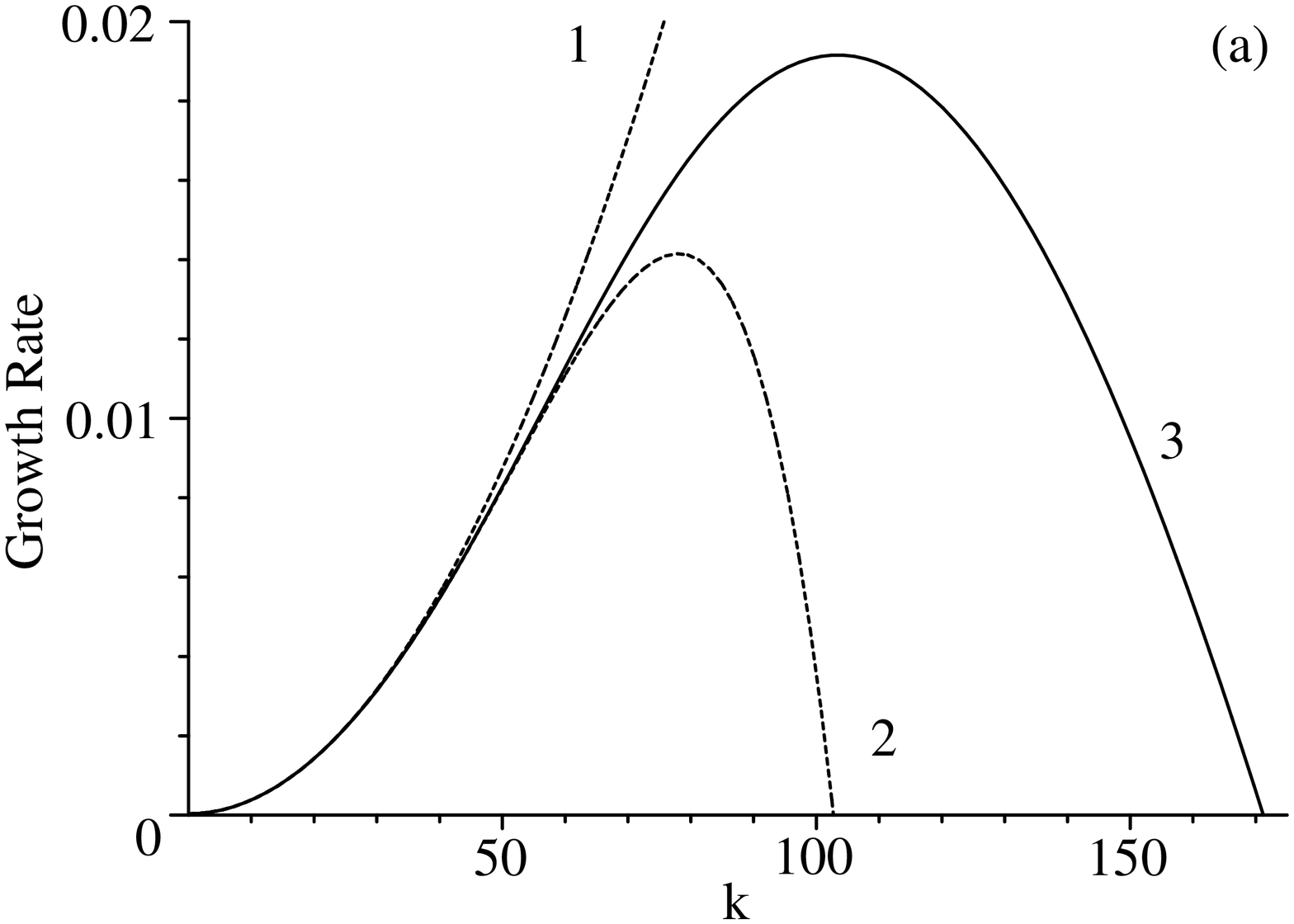}
\centering \includegraphics[width=2.3in,angle=-90]{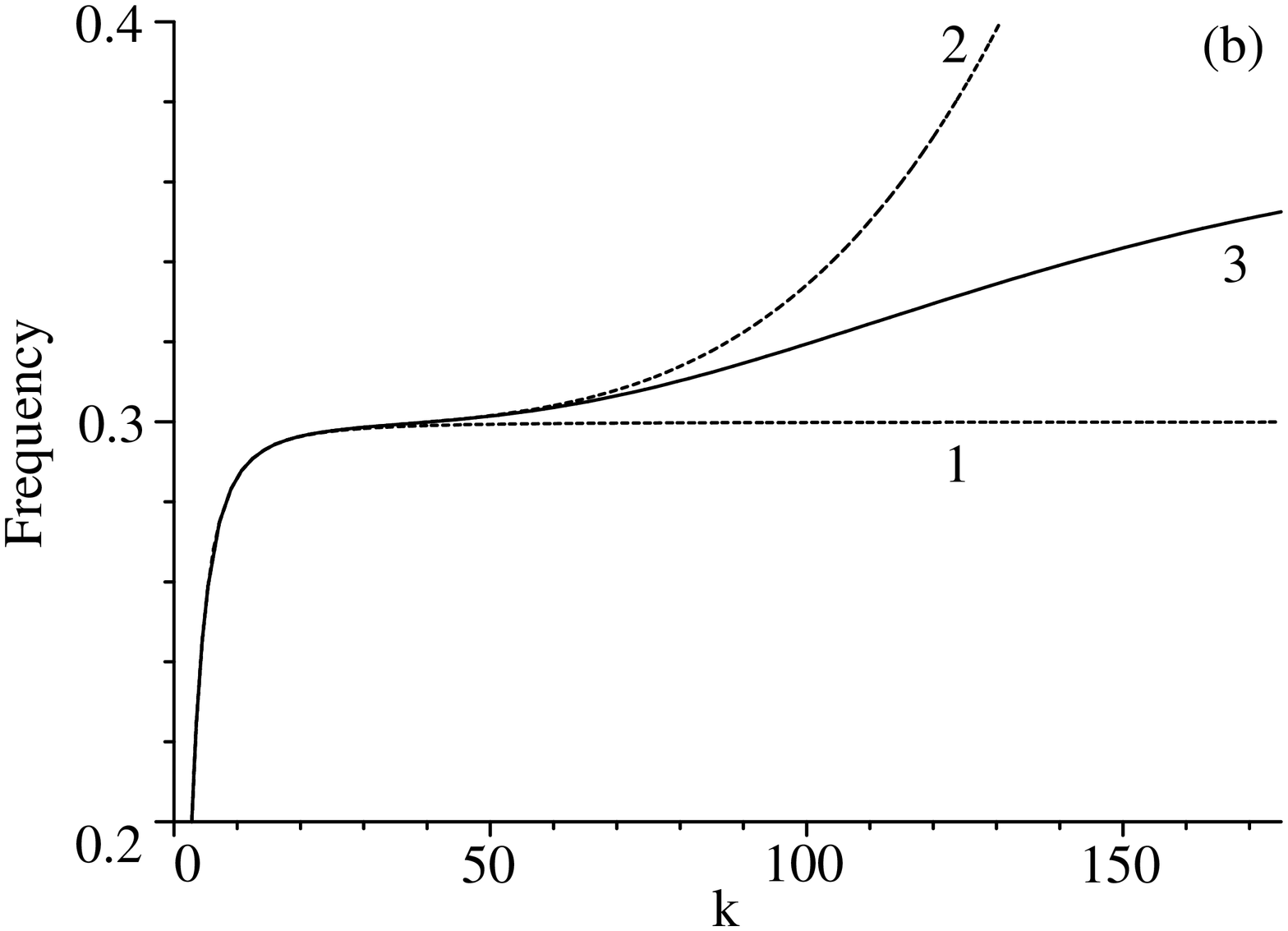}
\caption{\label{fig1} Grows rate ${\rm Re}(\lambda(k))$ (a), and
frequency ${\rm Im}(\lambda(k))$ (b) of travelling waves
(\ref{ncm2}) for the first convective mode. Here are:
$\varepsilon=0.00153, \sigma=7, \tau=1/81$ and $N=0.3$. Curves 1
are one term approximations; curves 2 are two term approximations;
curves 3 are exact solutions of the equation (\ref{dis2}).}
\end{figure}
where constants are: $C_1=1+\tau+\sigma$ and
$C_2=\tau+\sigma+\tau\sigma$. It is interesting to note, that when
$N_{*}>N>0$ functions $F_1,F_2,F_3$ are positive. The growth rate,
caused by thermohaline convective instability can be written as
follows~\cite{K7}:
\begin{equation}\label{INC} % (6)
{\rm Re}(\lambda) = \varepsilon^2\varkappa^2 F_1-\varepsilon^6
\frac{\varkappa^8}{k^2 N^2} F_3+\cdots\,\,\,.
\end{equation}
One can see (FIG. \ref{fig1}), that for a given mode with number
$n$, the growth rate is maximal for some $k$, which determines
horizontal size of the most prominent convective cells. Also the
first convective mode has maximal growth rate, so that convective
cells tend to be tall and thin. Really about 30 first modes for
$N=0.3$ have positive growth rate (FIG. \ref{figa}), but the most
prominent cells any way will be tall and thin, as having
relatively more large rate of growth. For simplicity further we
consider only the first convective mode.
\begin{figure}[ptb]
\centering \includegraphics[width=2.3in,angle=-90]{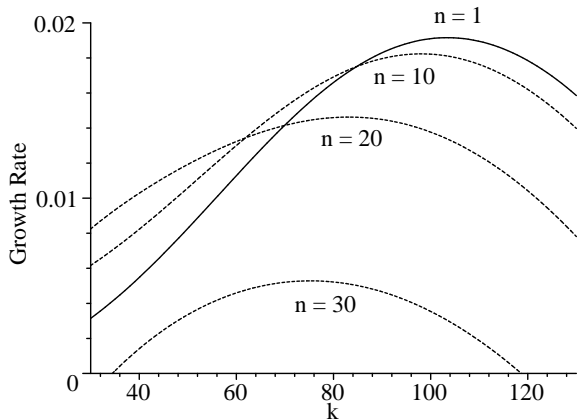}
\caption{\label{figa} Grows rate ${\rm Re}(\lambda(k,n))$ for
convective modes with different mode number $n$. All parameters are
the same as in the FIG. \ref{fig1}}
\end{figure}

Although above developed perturbation approach gives qualitatively
true estimates, for more accurate results one should immediately
solve algebraic equation (\ref{dis2}). Rewrite it in another form,
with new introduced variables
$P^2=\varepsilon^2\varkappa^2\approx\varepsilon^2 k^2$,
$X=\lambda/P^2$ and $Y=N^2/P^4$.
\[(X+\sigma)(X+1)(X+\tau)+Y(X+\gamma)=0.\]
Roots of this equation depend on parameter $Y$, so that finally
$\lambda$ depends on horizontal wave number $P$.

Consider actual oceanographical system such as an inversion of
thermohaline staircase. Let it has thickness $h=250$ cm and
temperature difference $\delta T=0.1^{\circ}C$, also $\sigma=7$ and
$\tau=1/81$. In this case $\varepsilon=1.53\times 10^{-3}$ and
non-dimensional critical buoyancy frequency $N_{*}=0.35136$. For
$N=0.2764$ the most unstable mode has $P_* = 0.1599579$ and width of
convective cell $l_c=\pi\varepsilon h/P_* \approx 7.7$~cm. For
comparison formula (\ref{INC}) gives $P_* = 0.126$, i.e. somewhat
less than the exact value. From (\ref{INC}) one can extract
dependence of $P_*$ from $N$, having form $P_* = [N^2 F_1/(3
F_3)]^{1/4}$.

\begin{figure}[tbh]
\centering \includegraphics[width=2.29in,angle=-90]{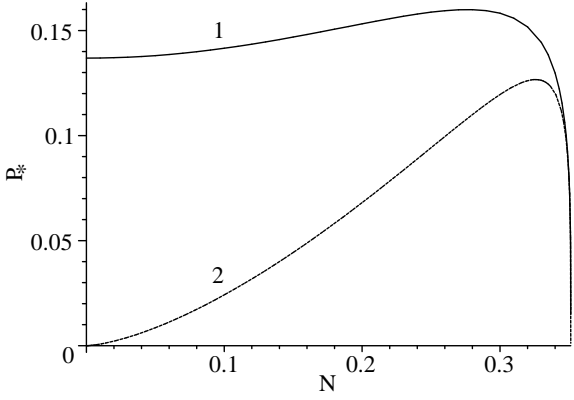}
\caption{\label{fig2}Dependence of the wave number $P_*
=k\varepsilon$ of the most unstable mode from buoyancy frequency
$N$. Curve 1 is the exact numerical solution; curve 2 is the
approximation by formula (\ref{INC}).}
\end{figure}

From picture (FIG. \ref{fig2}) one can easily see, that when value
of $N$ becomes slightly small than its critical value $N_* \approx
0.35136$, value of $P_*$ abruptly (as $(N_* - N)^{1/4}$) increases
and becomes maximal at $N\approx 0.3$. When $N$ becomes even more
less $P_*$ decreases to $P_* \approx 0.13688$ for $N = 0$. It should
be emphasized that $P_*$ is nearly independent of $N$ in considered
case.

This result for thermohaline convection at large $R_S$ is
sufficiently different from that for small $R_S$, when critical
wave number is $k_*=\pi/\sqrt{2}$~\cite{Huppert:1976a}. For our
case typical wave numbers are of the order $0.1/\varepsilon$. This
estimate is more accurate, than mentioned in~\cite{Turner:1973}.
Thus convective cells for large $R_S$ have tall and thin geometry
from linear stability analysis. Physically one can understand this
effect from consideration that when Rayleigh numbers are large,
the buoyancy forces acting vertically are also large in comparison
with forces of inertia of liquid particles determining width of
cells.

\section{FINE STRUCTURE GENERATION AND DIFFUSIVE CHAOS}\label{sec3}

Now we will study nonlinear vertical modulations of amplitude of
our tall thin convective cells.

At first, introduce new small parameter, extracted from geometry
of the convective cells: $e = l_c/h = \pi\varepsilon/P_* \approx
20\varepsilon$. Rescale variables $\psi=e^2\psi'$,
$\theta=e\theta'$, $\xi=e\xi'$ (prime will be omitted), and
introduce one more small parameter $E=\varepsilon/e\approx 1/20$.
After changing of the space scale from $h$ to $l_c$ basic system
(\ref{maineq2}) appears in the form:
\begin{eqnarray} \label{maineq3}
&&(\partial_t - \sigma E^2\Delta)\Delta\psi % \nonumber \\
%&&\qquad\qquad
+(\partial_x \xi - (1-N^2)
\partial_x\theta) = J(\Delta\psi,\psi),
 \nonumber \\
&&(\partial_t - E^2\Delta)\theta - \partial_x\psi =
J(\theta,\psi), \\
&&(\partial_t - \tau E^2\Delta)\xi - \partial_x
 \psi = J(\xi,\psi). \nonumber
\end{eqnarray}
At second, introduce slow vertical variable $Z=ez$ and slow time
$T=e^2 t$. In accordance with multi-scale decomposition technique
we get prolonged derivatives
\cite{Dodd:1982,Nayfeh:1976,Nayfeh:1980}: %4-6:
\begin{eqnarray*}
\partial_z &\rightarrow & e\partial_Z, \\
\partial_t &\rightarrow & \partial_t+e^2\partial_T, \\
\Delta &\rightarrow & \partial_x^2+e^2\partial_Z^2, \\
\Delta^2 &\rightarrow & \partial_x^4+2e^2\partial_x^2\partial_Z^2+
e^4\partial_Z^4, \\
\partial_t\Delta &\rightarrow & \partial_t\partial_x^2+
e^2\partial_T\partial_x^2+e^2\partial_t\partial_Z^2.
\end{eqnarray*}
Let buoyancy frequency somewhat less than its critical value
$N^2=N^2_* - e^2 R$. Parameter $R$ here is a forcing of the
system.  Equations (\ref{maineq3}) now get a form:
\begin{eqnarray} \label{maineq4} % (8)
&&(\partial_t - \sigma E^2\partial_x^2)\partial_x^2\psi +
(\partial_x \xi - (1-N^2_*) \partial_x\theta) = \nonumber \\
&&\qquad\qquad
- eJ_Z(\psi,\partial_x^2\psi) -
e^2[(\partial_T\partial_x^2+\partial_t\partial_Z^2 \nonumber \\
&&\qquad\qquad - 2\sigma E^2\partial_x^2\partial_Z^2)\psi
- R\partial_x\theta], \nonumber \\
&&(\partial_t - E^2\partial_x^2)\theta - \partial_x\psi =\nonumber \\
&&\qquad\qquad
- eJ_Z(\psi,\theta) - e^2(\partial_T - E^2\partial_Z^2)\theta, \\
&&(\partial_t - \tau E^2\partial_x^2)\xi - \partial_x\psi =\nonumber \\
&&\qquad\qquad - eJ_Z(\psi,\xi) - e^2(\partial_T - \tau
E^2\partial_Z^2)\xi. \nonumber
\end{eqnarray}
Solutions of these equations we will find as the asymptotic sets
by powers of the small parameter $e$:
\begin{eqnarray}\label{sets} % (9)
\psi &=& e\psi_1+e^2\psi_2+e^3\psi_3+\cdots \nonumber \\
\theta &=& e\theta_1+e^2\theta_2+e^3\theta_3+\cdots \\
\xi &=& e\xi_1+e^2\xi_2+e^3\xi_3+\cdots\,\,\, . \nonumber
\end{eqnarray}
After substitution of these expressions into equations (\ref{maineq4})
collect terms at the same powers of $e$. As a result we have
systems of equations for determining of the terms of the sets
(\ref{sets}). Thus, at $e^1$ we have following system:
\begin{eqnarray}\label{eq_e1} % 10
& & (\partial_t - \sigma E^2\partial_x^2)\partial_x^2\psi_1 +
(\partial_x\xi_1 - (1-N^2_*) \partial_x\theta_1) = 0, \nonumber \\
& & (\partial_t - E^2\partial_x^2)\theta_1 - \partial_x\psi_1 = 0, \\
& & (\partial_t - \tau E^2\partial_x^2)\xi_1 - \partial_x\psi_1 = 0. \nonumber
\end{eqnarray}
Choose for this system solution in the form of normal convective
mode travelling to the right, with constants of integration
$B(T,Z)$ and $C(T,Z)$, depending on slow variables.
\begin{eqnarray}\label{anz} % 11
\psi_1 &=& A(T,Z)e^{i\omega t-iKx} + c.c. \nonumber \\
\theta_1 &=& a_T(T,Z)e^{i\omega t-iKx}+B(T,Z)+ c.c. \\
\xi_1 &=& a_S(T,Z)e^{i\omega t-iKx}+ C(T,Z)+ c.c. \nonumber
\end{eqnarray}
Here wave number $K = K_*(N)$ is a horizontal wave number,
corresponding to the most unstable waves of convection, and
maximal value of $K=\pi$ from choice of the space scale, related
with convective cells. It is attained when $N\approx 0.3$.
Parameters of the normal mode (\ref{anz}) are related as follows:
\begin{eqnarray}\label{disa}
&& a_T=-\frac{iK}{i\omega+E^2 K^2}A, \qquad
a_S=-\frac{iK}{i\omega+\tau E^2 K^2}A, \nonumber \\
&& (i\omega+\sigma E^2 K^2)(i\omega+E^2 K^2)(i\omega+\tau E^2
K^2) \nonumber \\
&&\qquad\qquad\qquad\qquad\qquad + N^2(i\omega+\gamma E^2 K^2)=0.
\nonumber
\end{eqnarray}
Last formula is actually the dispersive relation
(\ref{dis2}), but for an infinitesimal vertical wave
number. Also for critical buoyancy frequency $N_*$ and wave
frequency $\omega$ we have:
\begin{eqnarray*}
N^2_* &=& \displaystyle{\frac{1-\tau}{1+\sigma}} -
(1+\tau)(\tau+\sigma)E^4 K^4, \\
\omega^2 &=& \displaystyle{\frac{1-\tau}{1+\sigma}}-\tau^2 E^4 K^4, \\
\omega^2 &=& N^2_* + (\sigma+\tau+\sigma\tau) E^4 K^4.
\end{eqnarray*}

System of equations at $e^2$ is the same as (\ref{eq_e1}) and does not lead
to any new results. System at $e^3$ is:
\begin{eqnarray*}\label{eq_e3}
&&(\partial_t - \sigma E^2\partial_x^2)\partial_x^2\psi_3  \\
&&\qquad +(\partial_x\xi_3 - (1-N^2_*)\partial_x\theta_3) =
-J_Z(\psi_1,\partial_x^2\psi_1) \\
&&\qquad
-(\partial_T\partial_x^2+\partial_t\partial_Z^2-2\sigma
E^2\partial_x^2\partial_Z^2)\psi_1+R\partial_x\theta_1, \\
&&(\partial_t - E^2\partial_x^2)\theta_3 \\
&&\qquad - \partial_x\psi_3 = - J_Z(\psi_1,\theta_1)
- (\partial_T - E^2\partial_Z^2)\theta_1,  \\
&&(\partial_t - \tau E^2\partial_x^2)\xi_3 \\
&&\qquad - \partial_x\psi_3 = - J_Z(\psi_1,\xi_1) - (\partial_T -
\tau E^2\partial_Z^2)\xi_1.
\end{eqnarray*}
After substitution into the right parts of these equations
expressions (\ref{anz}) we get a system with resonating right parts,
breaking regularity of the asymptotic expansions (\ref{sets}).
The condition of the absence of secular terms in this case takes
form of so called ABC system \cite{Balmforth:1998} % 7
(intermediate calculations are omitted):
\begin{eqnarray}\label{abc} % 12
\partial_T A &=& E^2\beta_1\partial_Z^2 A+R\beta_2 A-
\beta_3 A\partial_Z B+\beta_4 A\partial_Z C, \nonumber \\
\partial_T B &=& E^2\partial_Z^2 B-E^2\beta_5\partial_Z|A|^2, \\
\partial_T C &=& \tau E^2\partial_Z^2 C - \tau
E^2\beta_6\partial_Z|A|^2. \nonumber
\end{eqnarray}
Here coefficients are:
\begin{eqnarray*}\label{koabc}
\beta_0 &=& 1+\frac{1}{i\omega+E^2 K^2}\left[(i\omega+\sigma E^2
K^2)\right. \\
&&\qquad\qquad\qquad\qquad\qquad\qquad \left. - \frac{(1-\tau)E^2
K^2}{(i\omega+\tau E^2 K^2)^2}\right], \\
\beta_1 &=& \left\{ \left( \frac{i\omega}{E^2 K^2}+2\sigma \right)
+\frac{1}{i\omega+E^2 K^2}\right. \\
&&\qquad \left. \times \left[(i\omega+\sigma E^2 K^2)+
\frac{(1-\tau)i\omega}{(i\omega+\tau E^2 K^2)^2} \right]
\right\}\beta_0^{-1}, \\
\beta_3 &=& \left[(i\omega+\sigma E^2 K^2)+
\frac{1}{i\omega+\tau E^2 K^2}\right]\beta_0^{-1}, \\
\beta_2 &=& \frac{\beta_0^{-1}}{i\omega+E^2 K^2} \qquad
\beta_4=\frac{\beta_0^{-1}}{i\omega+\tau E^2 K^2}, \\
\beta_5 &=& \frac{2 K^4}{\omega^2+E^4 K^4}, \qquad \beta_6=\frac{2
K^4}{\omega^2+\tau^2 E^4 K^4}.
\end{eqnarray*}
Thus in this article we have derived ABC system of amplitude
equations for travelling waves of double-diffusive convection in a
limit of high Hopf frequency (large $R_S$) in the infinite
horizontal layer.

Equations (\ref{abc}) have nontrivial solutions, describing such
phenomena as diffusive chaos \cite{Ahr} and can be used for
simulation of formation of patterns, like vertical fine structure
of temperature and salinity in some areas of the ocean, for
instance, in inversions of thermohaline staircases.

Transform system (\ref{abc}) to more convenient form by
introducing new time variable $T'=E^2 T$, and applying following
substitutions:
\begin{eqnarray}%\label{koabc}
& A'=A E^{-1}\sqrt{\beta_5|\beta_3|}\exp(-iR\beta_{2R}T'E^{-2}) & \nonumber \\
& B'=|\beta_3|E^{-2}B \qquad C'=|\beta_4|E^{-2}C. & \nonumber
\end{eqnarray}
System (\ref{abc}) now gets form (primes are omitted):
\begin{eqnarray}\label{abc2} % (13)
\partial_T A &=& \beta_1\partial_Z^2 A+\alpha_2 R A-
\alpha_3 A \partial_Z B+\alpha_4 A \partial_Z C, \nonumber \\
\partial_T B &=& \partial_Z^2 B-\partial_Z |A|^2, \\
\partial_T C &=& \tau \partial_Z^2 C-\tau\alpha_6\partial_Z |A|^2.
\nonumber
\end{eqnarray}
Here are coefficients: $\alpha_2=\beta_{2R}E^{-2}$,
$\alpha_3=\beta_3/|\beta_3|$, $\alpha_4=\beta_4/|\beta_4|$
and $\alpha_6=\beta_6|\beta_4|/(\beta_5|\beta_3|)$.

We developed numerical models for parallel calculations of system
(\ref{abc2}) based on explicit and Dufort-Frankel schemes. For
numerical experiments it were chosen parameters $\sigma=7$,
$K=\pi$, $E=0.05$, and two values of Lewis number: $\tau=1/10$ and
$\tau=1/81$. Governing parameter $R$ was in the range from $0.1$
till $50$. Number of vertical dots $n$ was from $256$ till $2048$
to resolve microstructure.
\begin{figure}[ptb]
\centering \includegraphics[width=2.3in,angle=-90]{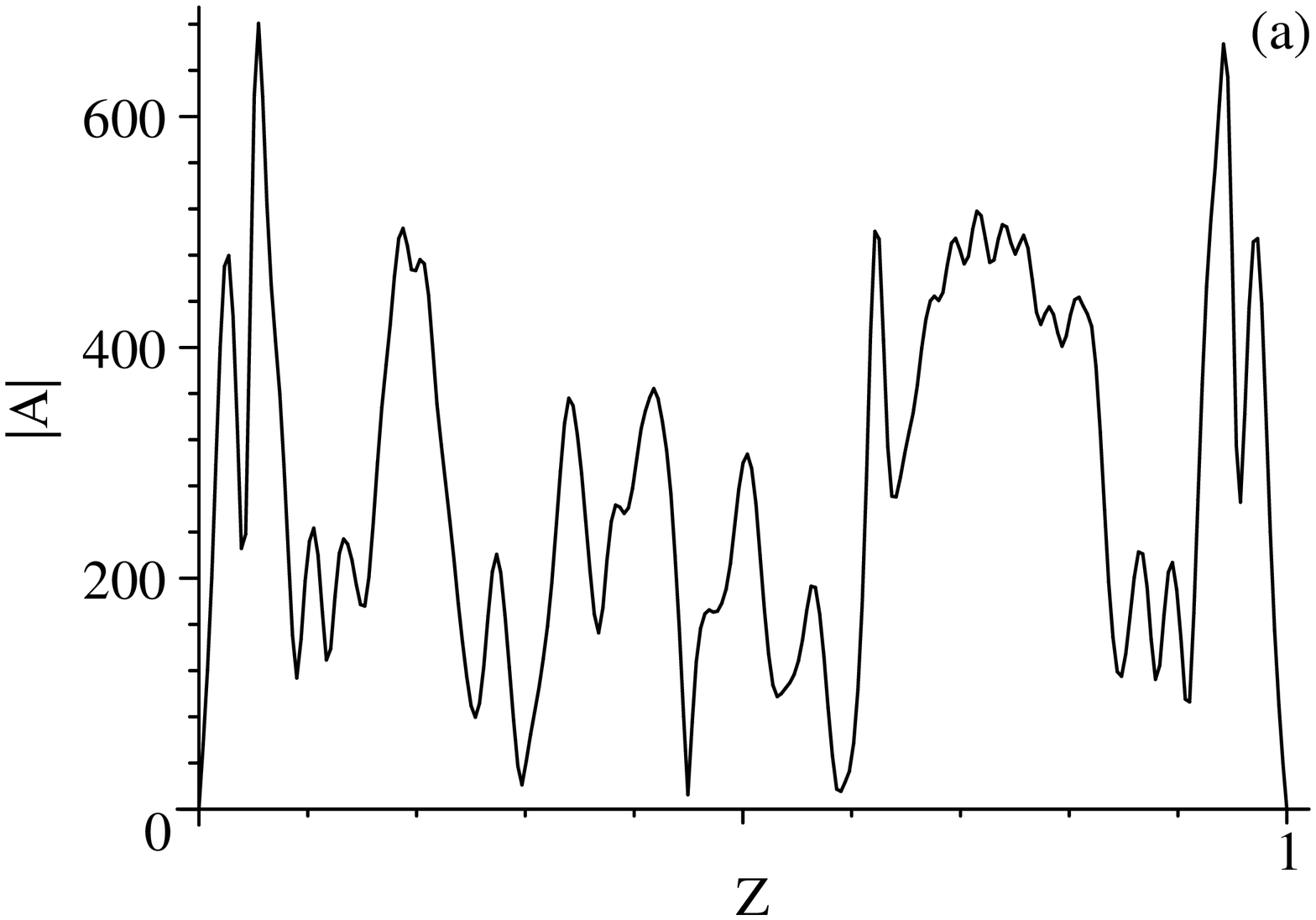}
\centering \includegraphics[width=2.3in,angle=-90]{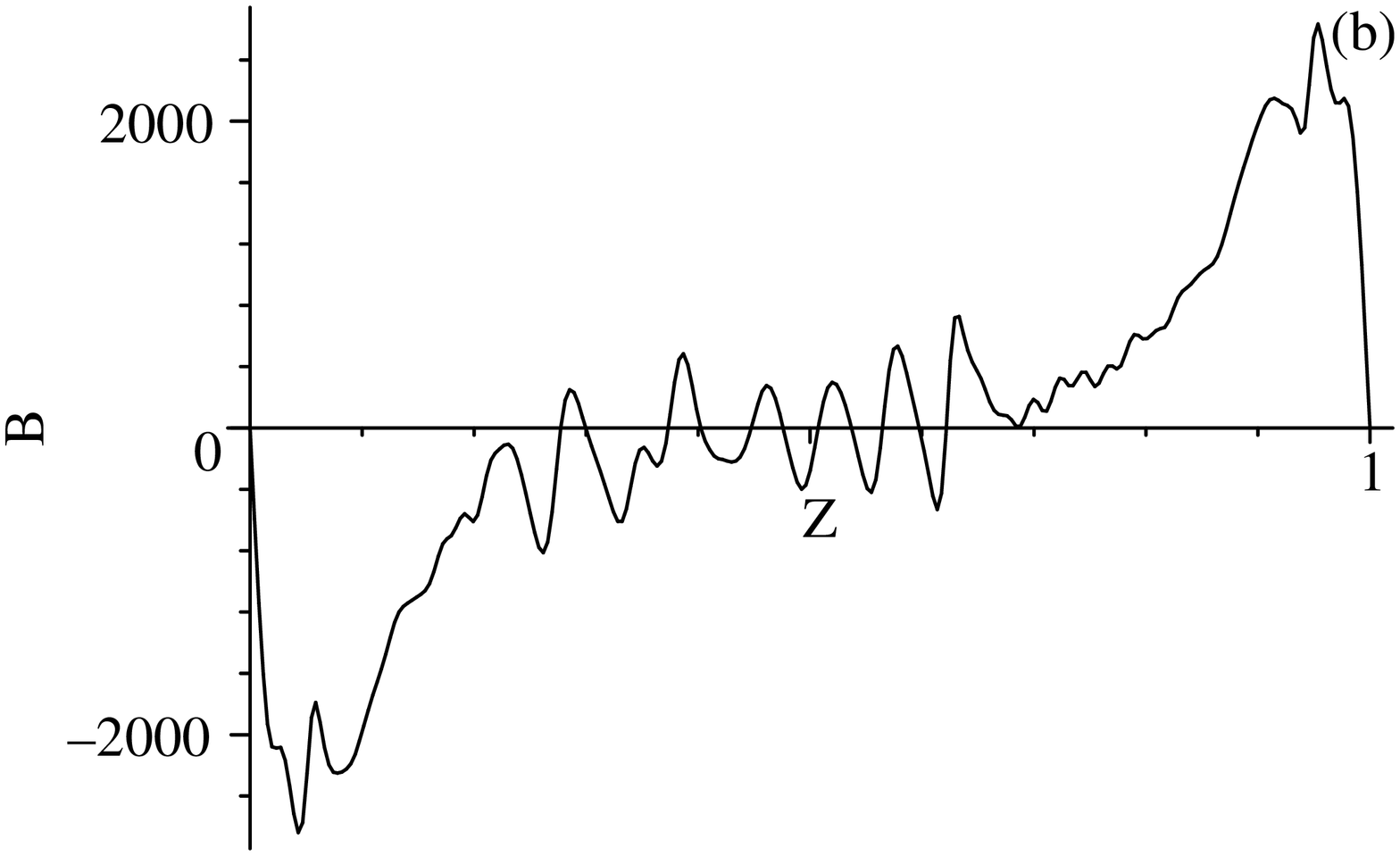}
\centering \includegraphics[width=2.3in,angle=-90]{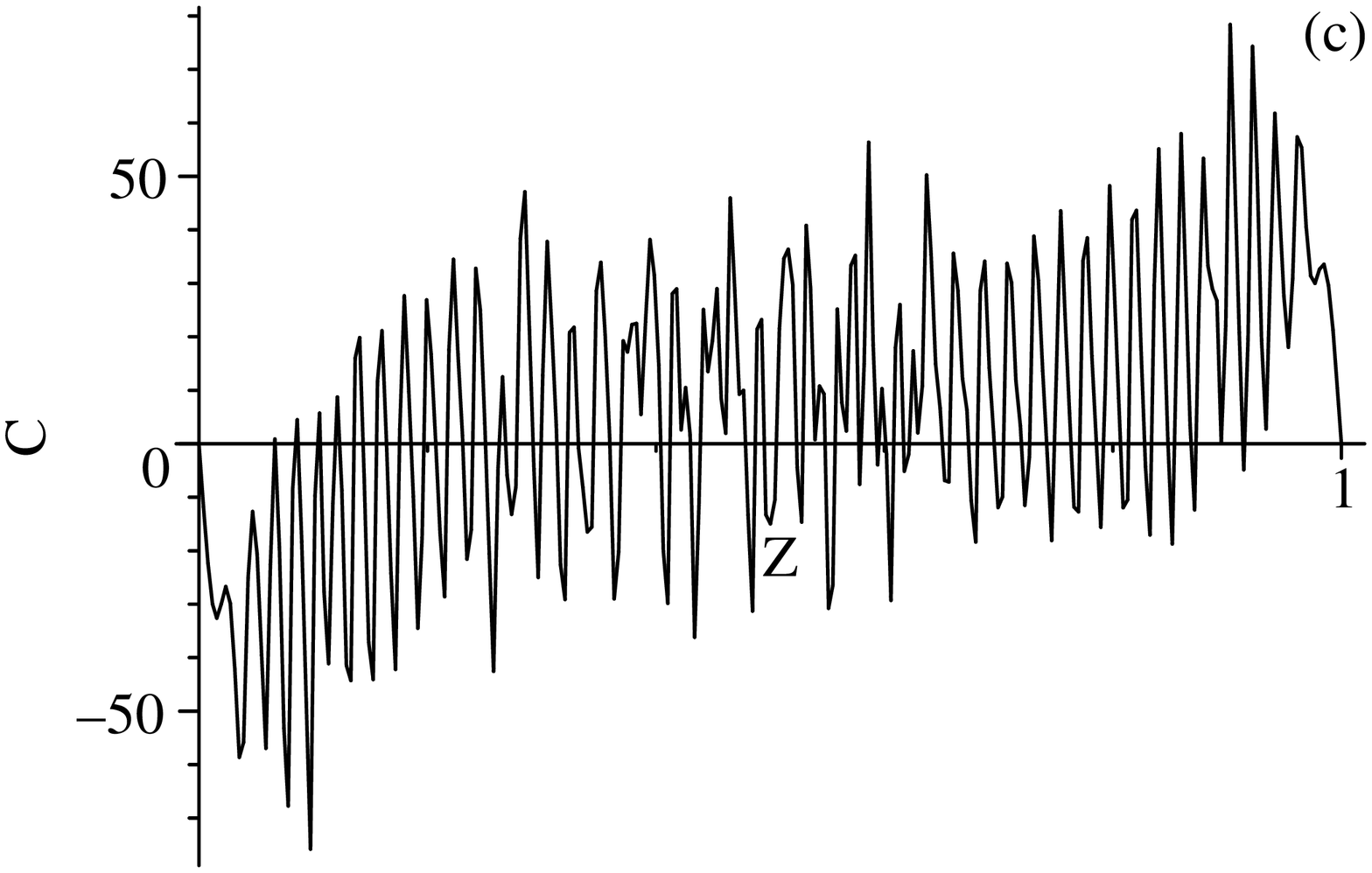}
\caption{\label{fig3} Numerical solution of the system (\ref{abc2})
at the time $t=9.28$ hours. Here variables: $|A(T,Z)|$ (a), $B(T,Z)$
(b), $C(T,Z)$ (c). Forcing parameter is $R=16$, also
$\varepsilon=0.00153, \sigma=7, \tau=1/81$. Layer depth $h=250$ cm,
the number of grid points is 2048. Dimensional variations in
temperature and salinity are proportional to $B$ and $C$, its can be
estimated as $6.9\cdot 10^{-5}\cdot B\cdot\delta T\, ^{\circ}C$ and
$6.0\cdot 10^{-5}\cdot C\cdot\delta S\, \%_o$ respectively.
Amplitude of stream function is proportional to $|A|$ and can be
estimated as $6.9\cdot 10^{-4}\cdot |A|\, cm^2/sec$. Maximal
variation in temperature, for instance, is about 20\% from $\delta
T$. }
\end{figure}

\begin{figure}[ptb]
\centering \includegraphics[width=2.3in,angle=-90]{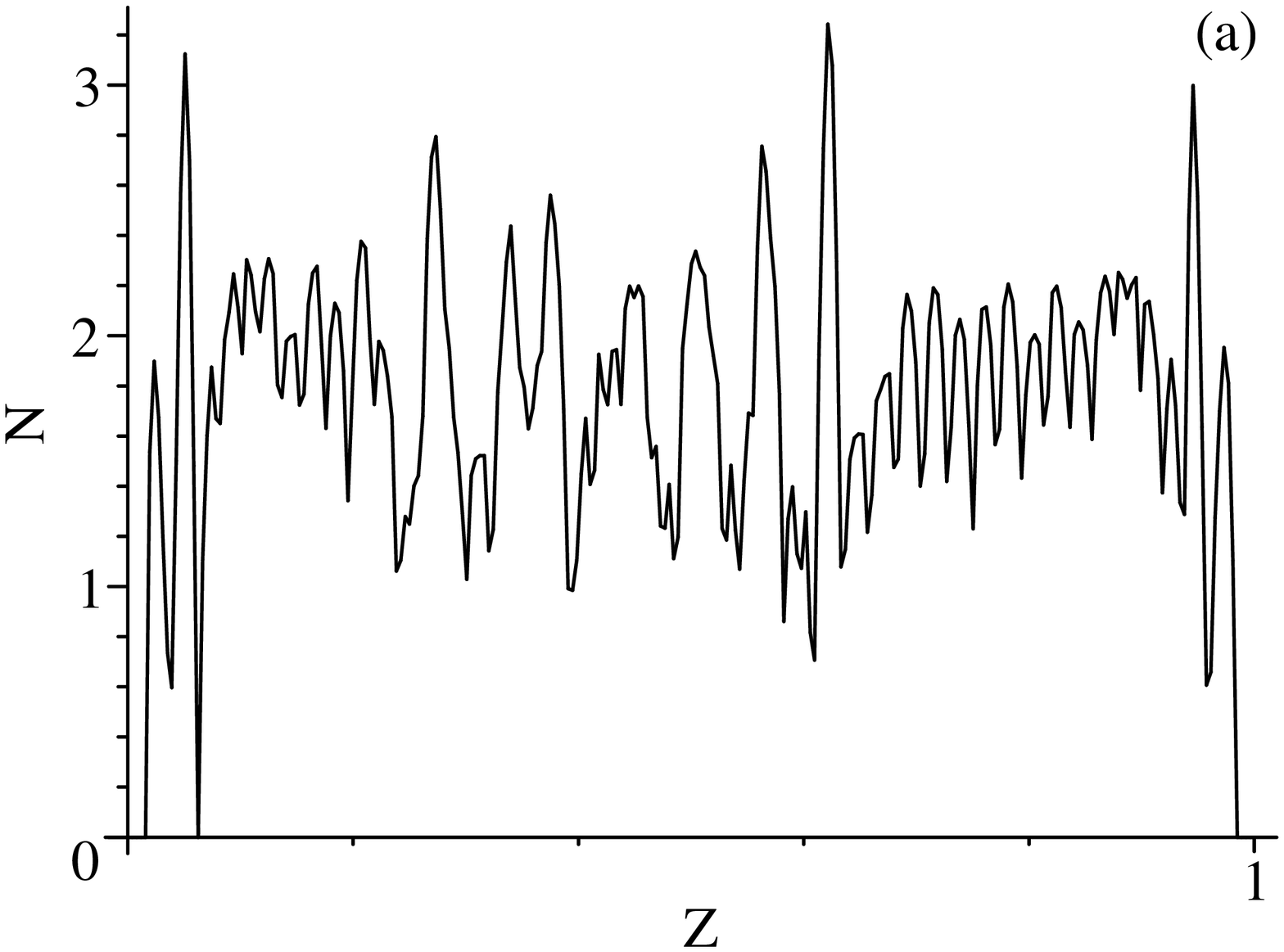}
\centering \includegraphics[width=2.3in,angle=-90]{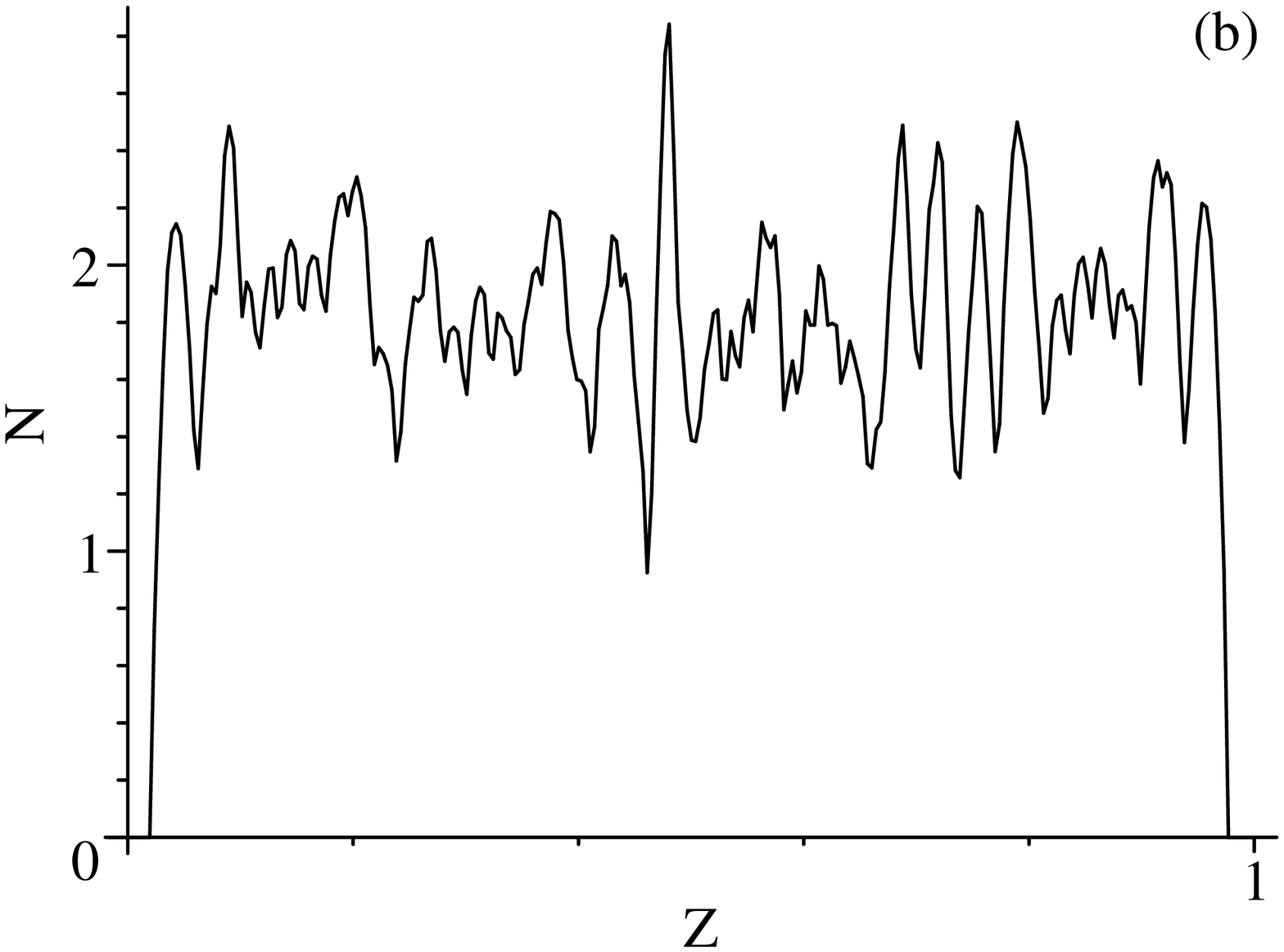}
\caption{\label{fig4} Buoyancy frequency $N$ (cycles per hour)
vertical microstructure for time $t=9.28$ (a), and $t=11.6$ (b).
Other parameters are the same as in the FIG. \ref{fig3}.}
\end{figure}

System (\ref{abc2}) was integrated numerically on multiprocessor
computer MVS-1000/16 with zero boundary conditions and sinusoidal
initial conditions for dependent variables. In the most cases the
initial state was destroyed after some time via a multiple Eckhaus
instability (birth of convective cells, \cite{Heske:1994}) and was
followed by diffusive chaos state, with strong space-time
irregularity. In this case mean profiles of temperature and
salinity become perturbed so that all layer of inversion becomes
divided on $10-30$ small layers (see FIG. \ref{fig3} for
$\tau=1/81$ and time $t=9.28$ hours). Buoyancy frequency (FIG.
\ref{fig4}) becomes very irregular, and all this fine structure
slowly changes with the time.

\section{CONCLUSION}

\begin{table}%[htb]
\caption{\label{tab1}Parameters estimations for inversion of
thermohaline staircase. For all cases it is true $T=15^{\circ}C$,
$S=36\%_o$, $\sigma=7$, $\tau=1/81$, $t_0$ -- diffusive time
scale, $t_0'$ -- main time scale, $N_0=(1-\tau)/(1+\sigma)$ --
limit of critical buoyancy frequency at $\varepsilon = 0$, $400\,
t_0'$ -- time of establishing of diffusive chaos in the
inversion.}
\begin{ruledtabular}
\begin{tabular}{llll}
Parameter &  1   &  2   &  3   \\
\hline
$h\, [cm]$   & 400.0  & 250.0  & 100.0 \\
$\delta T\, [^{\circ}C] $ & 1.0  & 0.1  & 0.1 \\
$\delta S\, [\%_o] $ & 0.33 & 0.033  & 0.033 \\
$R_T$   & $9.5 \times 10^{11}$  & $2.3 \times 10^{10}$
        & $1.48 \times 10^{9}$ \\
$R_S$   & $1.08 \times 10^{12}$  & $2.6 \times 10^{10}$
        & $1.69 \times 10^{9}$ \\
$\varepsilon$ & 0.0006 & 0.0015 & 0.003 \\
$e$     & 0.012 & 0.03 & 0.06 \\
$l_c\, [cm]$   & 4.7   & 7.7   & 6.0  \\
$t_0\, [sec]$  & $1.12 \times 10^{8}$  & $4.38 \times 10^{7}$
             & $7.0 \times 10^{6}$ \\
$t_0'\, [min]$ & 0.68  & 1.7  & 1.07 \\
$N_0\, [cyc/hr]$  & 4.95  & 1.97  & 3.13 \\
$400\, t_0'\,[hr]$ & 4.5 & 11.4 & 7.14 \\
\end{tabular}
\end{ruledtabular}
\end{table}

In this article we developed mathematical model, describing
formation of vertical convective patterns in two-dimensional
double-diffusive convection in a limit of high Hopf frequency for
an infinite horizontal layer.
A physical system, corresponding to such model is inversion of
thermohaline staircase. Some typical parameters of inversions are
presented in the table~\ref{tab1}.
It is known~\cite{Marmorino:1990} that parameters of
stratification in the inversions are often near the onset of
convection.
Also vertical microstructure (usually step-like) are often
observed in the inversions along with small scale
turbulence~\cite{Fedorov}.
Results of this work are in qualitative agreement with these
observations. For more comparison see also~\cite{Kerstein:1999}
and references therein.

Although we aimed to construct mathematical model with fine
structure generation in double-diffusive system without its
detailed relation with experimental data, we should note a few
points of such relation:
\begin{itemize}
\item Developed model predicts that fine structure should exist in given
system for a wide range of parameters with typical time of pattern
formation of about a few hours.
\item Fine structure has very irregular shape, slowly changing with the
time in accordance with solution of ABC-system of diffusive chaos
type. Mean profiles of temperature and salinity become perturbed
so that all layer of inversion becomes divided on $10-30$ small
layers.
\item Buoyancy frequency structures have peak emissions with amplitude
and width with reasonable agreement with observational
data~\cite{Fedorov} in the cases when temperature and salinity
differences per layer are relatively small, as in the
table~\ref{tab1}.
\end{itemize}

Of course, developed model, based on ABC-system of amplitude
equations has its limitation of weakly nonlinear approximation,
which hardly allows to get "full-fledged" step-like vertical
structure of density, as noted in~\cite{Balmforth:1998}. Partially
this defect can be overcome by regarding amplitude equations
arising at higher orders of small parameter in multi-scale
decomposition method. But this is a matter of further works along
with more detailed comparison of predicted fine structure
parameters with experimental and observational data.

\begin{acknowledgments}
This work is supported by the Governmental Contract No.
10002-251/$\Pi$-17/026-387/190504-301.
\end{acknowledgments}


\begin{thebibliography}{24}

\bibitem{Turner:1973} J. S. Turner, {\it Bouyancy Effects
in Fluids} (Cambridge University Press, 1973).

\bibitem{Grimshaw:1979} R. Grimshaw,
Phil. Trans. R. Soc. Lond. A, {\bf 292}, 391 (1979).

\bibitem{Grimshaw:1982} R. Grimshaw,
J. Fluid Mech., {\bf 115}, 347 (1982).

\bibitem{Vor} A. G. Voronovich, A. I. Leonov,
Ju. Z. Miropolskiy,
Oceanology (in russian), {\bf 11}, 490 (1976).

\bibitem{Kerstein:1999} A. R. Kerstein,
Dynamics of Atmospheres and Oceans, {\bf 30}, 25 (1999).

\bibitem{Balmforth:1998} N. J. Balmforth and J. A. Biello,
J. Fluid Mech. {\bf 375}, 203 (1998).

\bibitem{Sorkin:2002} A. Sorkin, V. Sorkin, I. Leizerson, Physica A.
{\bf 303}, 13 (2002).

\bibitem{Turner:1974} J. S. Turner, Ann. Rev. Fluid Mech.
{\bf 6}, 37 (1974).

\bibitem{Knobloch:1986} E. Knobloch, D. R. Moore, J. Toomre and
N. O. Weiss, J. Fluid Mech. {\bf 166}, 409 (1986).

\bibitem{K10} S. B. Kozitskiy, Journal of Applied Mechanics and
Technical Physics, {\bf 41}, 429 (2000).

\bibitem{K7} S. B. Kozitskiy, Linear stability problem for a system
with thermohaline convection in a limit of high Hopf frequency,
{\it Abstracts of Tenth Annual Meeting PICES} (Victoria, B.C., Kanada, 2001), P.~191.

\bibitem{Huppert:1976a} H. E. Huppert and D. R. Moore,
J. Fluid Mech. {\bf 78}, 821 (1976).

\bibitem{Dodd:1982} R. K. Dodd, J. C. Eilbeck, J. D. Gibbon and
H. C. Morris, {\it Solitons and Nonlinear Wave Equations}
(Academic Press Inc., London Ltd., 1982).

\bibitem{Nayfeh:1976} A. H. Nayfeh, {\it Perturbation
methods} (John Wiley and Sons, New York, London, Sydney, Toronto,
1973).

\bibitem{Nayfeh:1980} A. H. Nayfeh, {\it Introduction to
perturbation techniques}, (John Wiley and Sons, New York,
Chichester, Brisbane, Toronto, 1981).

\bibitem{Ahr} T. S. Akhromeeva, S. P. Kurdyumov, G. G. Malinetskiy
and A. A. Samarskiy {\it Chaos and dissipative structures in
reaction-diffusion systems}, (Nauka, Moscow, 1992).

\bibitem{Heske:1994} M. van Hecke, P. C. Hohenberg and W. van Saarloos,
Amplitude equations for pattern forming systems,
{\it Fundamental Problems in Statistical Mechanics VIII},
H. van Beijeren and M. H. Ernst, eds. (North-Holland, Amsterdam, 1994),
P.~245-278.

\bibitem{Marmorino:1990} G. O. Marmorino, Deep-Sea Res.
{\bf 38}, 1431 (1991).

\bibitem{Fedorov} K. N. Fedorov,
{\it The thermohaline finestructure of the ocean},
English edition, 1977, translated by D. A. Brown,
technical editor, J. S. Turner, (Pergamon, 1976).


\end{thebibliography}
\end{document}